# Developments in high-density Cobra fiber positioners for the Subaru Telescope's Prime Focus Spectrometer

Charles D. Fisher*[a], David F. Braun[a], Joel V. Kaluzny[a], Michael D. Seiffert[a], Richard G. Dekany[b], Richard S. Ellis[c], Roger M. Smith[c]

[a]Jet Propulsion Laboratory/California Institute of Technology, 4800 Oak Grove Dr., Pasadena, CA 91109 USA; [b]Caltech Optical Observatories, 1201 East California Blvd., Pasadena, CA 91125 USA; [c]California Institute of Technology, 1201 E. California Blvd., Pasadena, CA 91125 USA

## ABSTRACT

The Prime Focus Spectrograph (PFS) is a fiber fed multi-object spectrometer for the Subaru Telescope that will conduct a variety of targeted surveys for studies of dark energy, galaxy evolution, and galactic archaeology. The key to the instrument is a high density array of fiber positioners placed at the prime focus of the Subaru Telescope. The system, nicknamed "Cobra", will be capable of rapidly reconfiguring the array of 2394 optical fibers to the image positions of astronomical targets in the focal plane with high accuracy. The system uses 2394 individual "SCARA robot" mechanisms that are 7.7mm in diameter and use 2 piezo-electric rotary motors to individually position each of the optical fibers within its patrol region. Testing demonstrates that the Cobra positioner can be moved to within 5μm of an astronomical target in 6 move iterations with a success rate of 95%. The Cobra system is a key aspect of PFS that will enable its unprecedented combination of high-multiplex factor and observing efficiency on the Subaru telescope. The requirements, design, and prototyping efforts for the fiber positioner system for the PFS are described here as are the plans for modular construction, assembly, integration, functional testing, and performance validation.

## 1. INTRODUCTION

The Prime Focus Spectrograph (PFS) is a ground based instrument being developed for the Subaru Telescope on Mauna Kea, Hawaii. An earlier concept for this instrument was called the Wide-Field Multi Object Spectrometer (WFMOS). The PFS has a hex shaped array of 2394 fiber positioners in a 1.38 degree field of view at the prime focus of the telescope. The hex shape allows for efficient tiling of the sky and room on the focal plane for fiducial fibers and guide cameras, which are shown in Figure 2. The positioners virtually simultaneously place –f/2.4 fibers on up to 2394 targets in the field. The light from each target is fed through ~50m of fiber to a spectrograph located off the telescope. The current estimate is it takes 62 seconds to move the telescope and reacquire the new field of targets with 15 minute exposures. This will allow for approximately 45 fields to be observed per night. This will enable large scale surveys aimed at understanding dark energy as part of the Subaru Measurements of Images and Redshifts (SuMIRe) program.

*charles.d.fisher@jpl.nasa.gov; phone 1 818 393-5067; fax 1 818 393-4860; jpl.nasa.gov

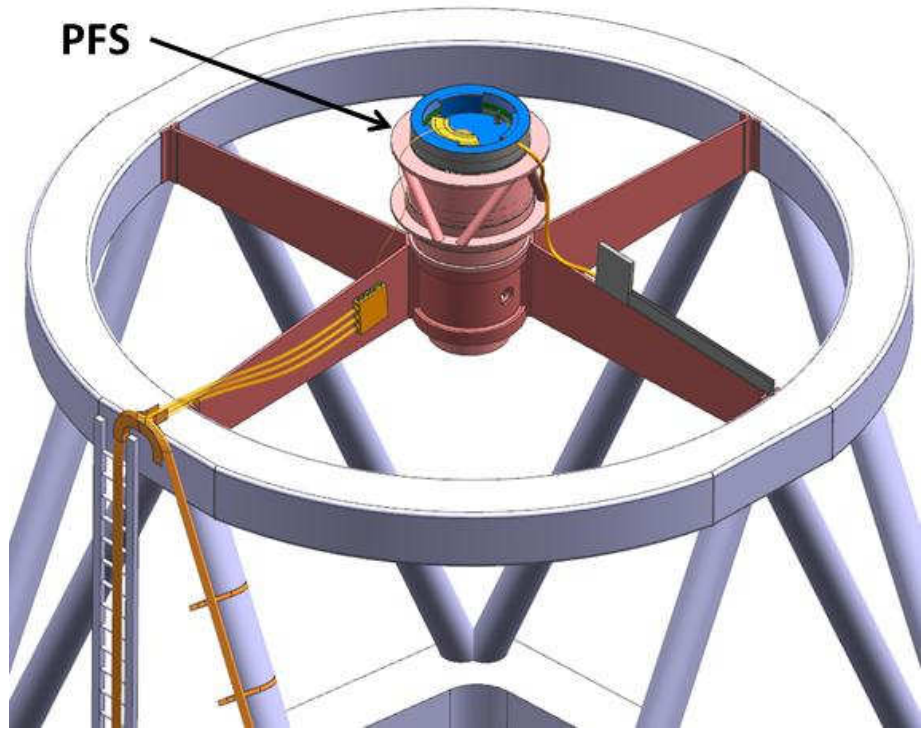

Figure 1. PFS on the Subaru Telescope

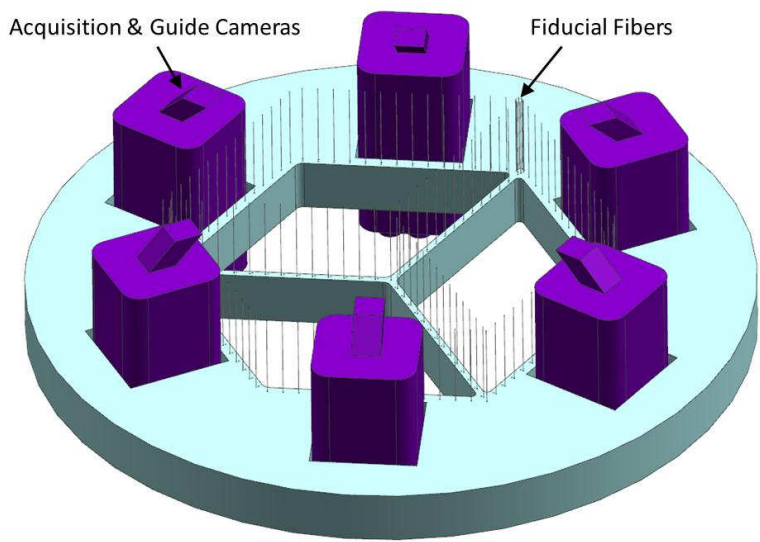

Figure 2. Guide camera and fiducial fiber locations on the focal plane (Cobra positioners not shown)

The Cobra fiber positioner was originally developed in 2008-09[1]. It is a theta-phi style mechanism containing two small piezo electric rotary motors developed specifically for the Cobra positioner[2]. The geometric and performance constraints are very similar for PFS as they were for WFMOS. The positioners are arranged in a close packed hex flat pattern with 8mm center-to-center spacing. The Cobra positioners have patrol regions that overlap with adjacent positioners to enable full coverage of the focal plane. The changes to the Cobra design for this development phase were related to the hard stops and to enhance the manufacturability of the piece parts as well as the assembly. These will be discussed in more detail in Section 3.

Table 1. Geometric Constraints

| Requirement | Value | Units |
|---|---|---|
| Positioner External Diameter | ≤ 7.7 | mm |
| 1st Stage Internal Diameter | ≥ 1.2 | mm |
| 2nd Stage Offset | 2.375 | mm |
| 1st Stage Hard Stop | ≥ 360 | deg |
| 2nd Stage Hard Stop | ≥ 180, < 360 | deg |

Table 2. Performance Constraints

| Requirement | 1st Stage | 2nd Stage | Units |
|---|---|---|---|
| Step Size | ≤ .084 | ≤ .167 | deg |
| Stall Torque | ≥ 346 | ≥ 337 | µN-m |
| Speed | ≥ 1.0 | ≥ 0.5 | rev/sec |

## 2. FOCAL PLANE CONFIGURATION

The focal plane was configured such that the high density of motors, drive electronics, and optic fibers could coexist at Subaru's Prime Focus and be serviceable while off the telescope on the Summit of Mauna Kea. With 4788 motors driven by ~15000 drive signals and 2394 optic fibers needing to be routed amongst them, a good configuration is crucial. The Cobra positioner close packed hex pattern was divided into 3 axisymmetric parallelograms (Figure 3), allowing the positioners to be grouped into 42 identical modules (Figure 4). The modularity greatly simplifies integration and verification of the positioners, as well as assembly and serviceability for the Prime Focus Instrument. Fiducial fibers are added to the focal plane to define the coordinate system for the Cobra positioners and monitor and account for distortions of the Wide Field Corrector when placing the fibers on their assigned targets. The fiducial fibers are located around the perimeter of the field and near the ends of the modules along the 3 adjacent boundaries of the parallelograms which are highlighted orange in Figure 5.

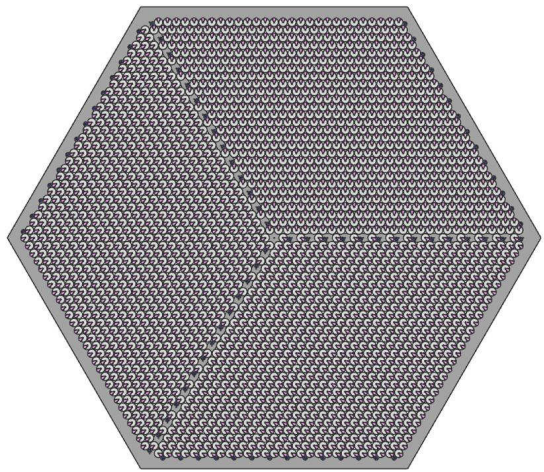

Figure 3. Focal plane layout of positioner modules.

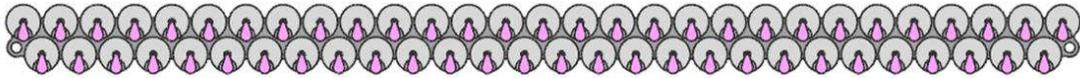

Figure 4. 57 Positioners arranged on a module.

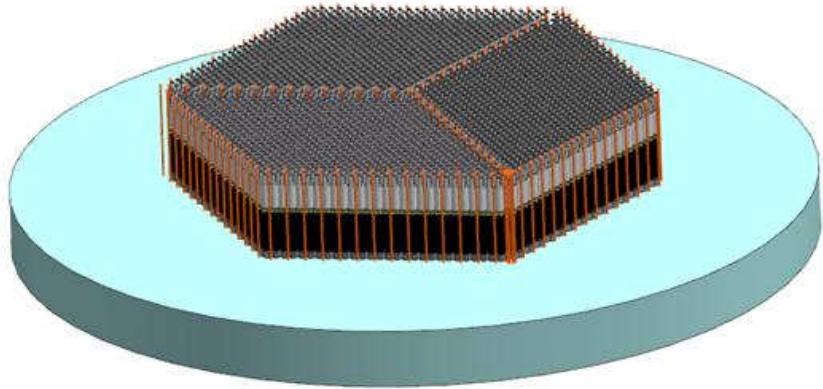

Figure 5. Fixed fiducial fiber locations

### 2.1 Cobra module

Each of the 42 Cobra modules contains 57 positioners mounted in two offset rows.  Each row has two drive electronics boards mounted to the back of the module allowing the electrical signals to be routed to the edge of the field away from the optic fibers.  The modularity enables testing prior to installation into the instrument.  The modularity concept was also extended to the fiber system allowing replacement of a single module should some part of it require servicing.

The current design has a fiber routed through the center of each Cobra to minimize the twisting and bending and to protect the fiber as it is routed through the instrument.  Routing the fiber through the Cobra places additional constraints on the design of the fiber routing.  In the event of future positioner servicing, a fiber routed through the center of the positioner would require cutting and re-polishing or replacing entirely. The module design also accommodates the fiber routing external to the Positioners which allow Positioner replacement without fiber rework.  Future testing will assess whether routing the fiber external to the Cobra positioner is acceptable for fiber protection and Focal Ratio Degradation due to the additional strains applied to the fibers.

## 3. DESIGN CHANGES

### 3.1 Hard stops

In the previous development cycle for the Cobra positioner the jamming of the floating hard stops were identified as a contributor to positioning errors and general erratic behavior.  The small floating hard stop on the second stage also had a propensity to fall out or become dislodged during basic handling of the Cobra resulting in hours of repair time.  These issues led to eliminating the >360° floating hard stop and replacing it with a static hard stop that would allow -20° to 200° motion on the second stage.  Having more than 180° of motion on the second stage ensures that the maximum patrol area can be reached.  The floating hard stop was still used on the first stage, but the aspect ratio was increased to reduce the likelihood that it would rotate and jam.

### 3.2 Design for manufacturing and assembly

Three sets of changes were also made to the Cobra design to enhance the manufacturability of the piece parts and minimize the assembly time in an effort to reduce the unit cost.  The first was to reduce the number of bonding operations required to build up the motors and the assembly.  Most of the bonded joints were replaced with set screws or were re-designed to be clamped in place.  The only bonds that still exist are the attachment of the piezo plates and ceramics to the motor stators and the mounting flange to the motor stator. The second change was to use a flex print cable for the second stage motor to keep the wiring within the 7.7mm allowable footprint. It also enabled a cable loop to be added to the same area where the 1st stage hard stop is located so that there is no motion of the electrical connectors near the base of the Cobra. The most important change in this effort was to more tightly control the surface finish and run out of the ceramic end caps on the motors. The intent is to increase the overall yield of the motors and minimize any need for hand polishing or refurbishment of the parts if the motors failed their screening performance tests. Ultimately these changes didn't significantly reduce the unit cost for the most recent build cycle of five Cobra's, but should lead to noticeable savings as production is scaled up to 2400 units.

## 4. PERFORMANCE TESTING

New Scale Technologies delivered five redesigned Cobra positioners to the Jet Propulsion Laboratory (JPL) in the spring of 2012 for performance testing. The goal of the performance testing was to evaluate the Cobra's ability to position a fiber within 5µm of a target and to determine how many move iterations were necessary to achieve this accuracy. Comparing the performance of the new design with the original Cobra would validate the design changes discussed earlier.

### 4.1 Test set up

JPL's performance testing test bed was designed to mimic the metrology setup that will be used on the PFS. As on the telescope, the fibers held by the Cobra positioners are backlit and imaged by a metrology camera to determine the location of each fiber. In the test bed, this is accomplished using a halogen lamp to illuminate the fibers, and a QSI 540i camera with a CCD size of 2048 x 2048 pixels to image the location of the fiber in pixel space. A translation stage was used to move the fiber a known distance and establish the pixel scale for the 7.4 µm/pixel, which relates pixel space to real space at the image plane at the fiber tips. The camera and lens were set up with a near 1:1 pixel scale ratio (actual pixel size to imaged pixel size), to insure a large enough image size on the CCD for accurate centroiding. With an illuminated fiber spot diameter of approximately 18 pixels (133µm), sub-micron centroid accuracy was easily achievable. This test bed setup allows for the accurate measurement of fiber position necessary to assess the performance of the new Cobra positioners.

### 4.2 Testing Overview

Each Cobra is first measured and characterized in the test bed prior to performance testing. The center of the 1st stage, the offset of the 2nd stage, and the length of the fiber arm must be determined in order to calculate the motor angles for any given fiber position. This information is measured by rotating each stage independently and taking multiple images of the backlit fiber around its arc of motion. A circle is fit to the centroid locations of each fiber image and the center location and radius is computed for each stage. Rotating the 1st stage gives its center, rotating the 2nd stage gives the fiber arm radius, and the distance between the centers of each stage is the offset between the 1st and 2nd stages. With these necessary measurements complete, the movement of each motor is characterized and stored in a motor calibration table. The motor calibration table also needs to track the average step size (degrees moved per motor step) of a motor as a function of its angular position for forward and reverse movements. The patrol area is divided into regions of 15° to 30° to accomodate variance in motor performance that may result from differences in fiber load, friction, etc. as the motor rotates. Such variance in performance can ruin positioning performance if it is not tracked and accounted for. Motor calibration tables are continuously updated during regular operation to keep up with any changes in motor performance over time. This method of calibration allows the software to accommodate a wider range of motor performance and helps reduce the number of iterations required to acquire a target by "learning" the subtle variances in each motor's performance.

With the measurement and characterization complete, each Cobra can begin the performance testing. The Cobra is first initialized to its home position by running the motors in reverse into the hard stops. The fiber is imaged, using Maxim DL camera control software, at the home position. Using MATLAB, the fiber's current position is calculated by centroiding the image of the backlit fiber. The Cobra's motor angles are calculated using the center, offset and radius values measured in the setup. When a new target is selected, the MATLAB software determines the number of moves required to move to the target based on the motor calibration tables. With this information, it generates a script that is used by the New Scale Technologies Squiggle motor control software to move the Cobra via the New Scale Technologies MC-1000 motor controller electronics. After the Cobra is repositioned, a new image is taken and compared to the target. Subsequent iterations are made until the centroid of the backlit fiber is within 5µm of the target position. This process, beginning with the initialization of the Cobra positioner, is repeated for each new target, and the number of iterations required to converge on a target is recorded as an assessment of the positioner performance. Targets were chosen throughout the patrol region, including stress cases with motor angles near their hard stops, or at the extremes of the patrol region to accurately portray actual usage.

### 4.3 Motor calibration comparison

The improved performance of the redesigned Cobra positioner was evident even prior to performance testing. When initially characterizing the motor performance, the average step size of the motors was seen to be highly consistent throughout the motion of the motors. The resulting motor calibration "maps" were uniform and showed that the motors lacked areas of erratic motion, which in the past made motor movement difficult to predict. Figure 6 shows a Stage 1 motor map for forward motion. In constructing this map, the Stage 1 motor was moved in increments of 200 steps and imaged after each movement. The uniformity of the image segments reveal the consistency of the step size throughout the motion. In testing prior Cobras (2009) the motor maps revealed areas of varying step sizes that plagued performance. The improvement in motor consistency is possibly a result of the redesigned hard stops. Eliminating the Stage 2 floating hard stop, and increasing the aspect ratio of the Stage 1 floating hard stop has no doubt reduced friction and/or jamming in the Cobras, making motion more consistent and predictable. Consistency of the motor performance improves accuracy of the moves and ultimately reduces the number of iterations required to position the fiber at a target.

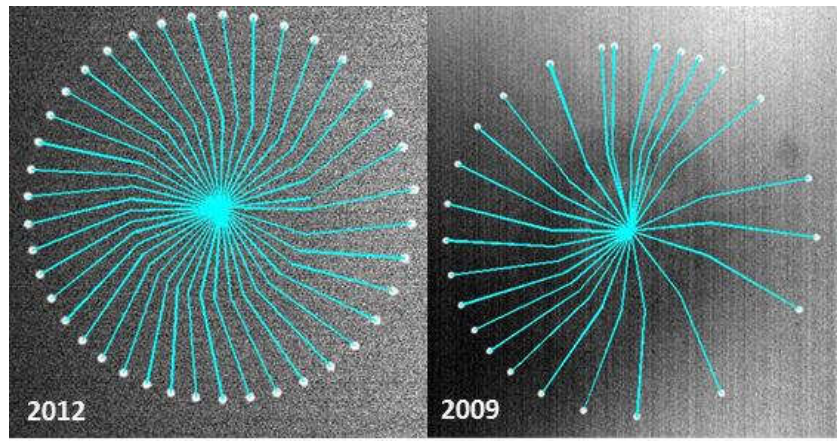

Figure 6. Motor Calibration Results

### 4.4 Comparison of performance

As suspected, the increased consistency of motion of the redesigned Cobra positioners translated to a marked increase in positioning performance. After four iterations, 78.4% of attempts by the redesigned Cobra fiber positioner had converged within 5μm of the target, compared to only 27.6% of the 2009 version. In comparing the plots in Figure 7 and 8 the 2012 iterations required to converge are reduced in most cases by at least 1 step, which is seen as a shift of the curve to the left in Figures 7 and 8. To achieve a 90% threshold of convergence, the new Cobra positioners needed six steps, whereas the 2009 version required an additional step. This increase in performance translates to less time spent repositioning fibers on the PFS, and more time for science measurements.

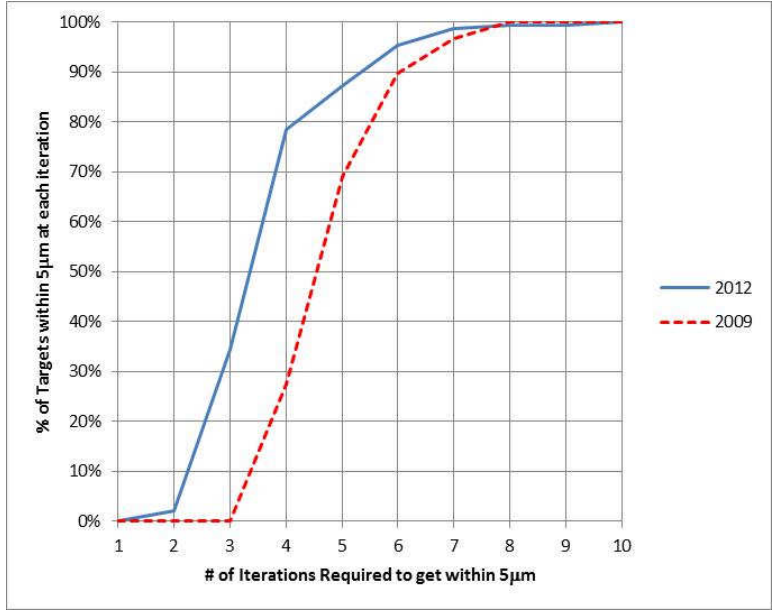

Figure 7. Comparison of total fibers on target by iteration.

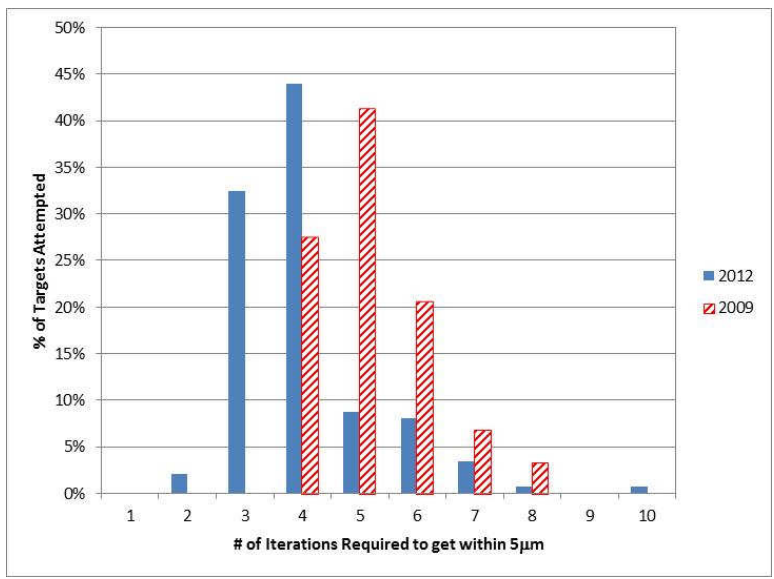

Figure 8. Comparison of distribution of iterations required to move to target.

The redesign of the Cobras resulted in an obvious improvement in performance. The new Cobra positioners are able to converge within 5 μm of a target in less iterations than before. There is still potential for further improvement. For example, the gains that resulted from removing the floating hard stop design from Stage 2 can also be imagined for Stage 1. Although limiting the motion of Stage 1 to less than 360 degrees, as required by a fixed hard stop design, will sacrifice a small amount of the patrol area it would potentially improve the consistency of the Stage 1 motor motion. This would improve the accuracy of Stage 1 and potentially reduce the iterations required to reach a target. Alternatively, additional work on the design of the Stage 1 floating hard stop may make it more consistent and even less likely to jam or cock.

# 5. PATH FORWARD

Over the next 12-18 months a number of development tests are planned to address aspects of Cobra positioner performance that have been untested. The following sections discuss a few of these tests.

## 5.1 Collision avoidance

A key feature of the Cobra design is that it offers overlapping patrol areas with the adjacent positioners so there is total coverage of the focal plane. That feature also creates the problem of colliding with adjacent positioners. The concept for avoiding collisions is to move the fibers to an intermediate waypoint for the first two move iterations, then have the final moves towards the target take the fiber in a radial direction. The actual implementation of this strategy and development of the details in the algorithms will occur with a test bed comprised of at least 5 functioning Cobras and a few dummy Cobras.

## 5.2 Running multiple Cobras simultaneously

All testing to date has only involved running a single Cobra positioner at any given time. Demonstrating the ability to run multiple Cobras at the same time is a key step in proving that the system of 2396 positioners is possible. Mechanically there aren't any challenges beyond the collision avoidance discussed earlier, but there are significant challenges with electronics and software.

The challenges for the electronics are related to miniaturization and packaging. To locate the motor drive electronics on each Cobra module requires the electronics to have a packaging height of less than 6mm. For software communication and data handling will be paramount.

Initially multiple Cobras will be run simultaneously using existing New Scale Technologies MC-1000 boards dedicated for each motor and all connected via RS-232 and USB. The end goal is to develop custom electronics that will run all motors on a module and communicate via Ethernet.

# 6. SUMMARY

A second generation Cobra positioner was designed based on lessons learned from the original prototype built in 2009. Improvements were made to the precision of the ceramic motor parts and hard stops were re-designed to minimize friction and prevent jamming. These changes resulted in reducing the number of move iterations required to position the optical fiber within 5μm of its target. There are still many tests still to be performed that will validate system level performance, but on an individual level the Cobra positioner demonstrates excellent performance and will enable the PFS instrument to make unprecedented measurements of the universe.

# 7. ACKNOWLEDGEMENTS

Part of this research was carried out at the Jet Propulsion Laboratory, California Institute of Technology, under a contract with the National Aeronautics and Space Administration. All of the R&D funding to date has provided by JPL.

Support was provided from the Funding Program for World-Leading Innovative R&D on Science and Technology (FIRST) "Subaru Measurements of Images and Redshifts (SuMIRe)", CSTP, Japan.

The authors would like to thank the following individuals who provided support in conducting this work: Justin Vacca (New Scale Technologies), Rob Culhane (New Scale Technologies), Conrad Hoffman (New Scale Technologies) and Dan Viggiano (New Scale Technologies).